\documentclass[aps,pre,preprint,superscriptaddress]{revtex4}
\usepackage{graphics,epsfig,wasysym}

\begin{document}

\centerline{\Large\bf Triezenberg-Zwanzig expression}
\centerline{\Large\bf for the surface tension of a liquid drop}
\vskip 10pt
\centerline{Edgar M. Blokhuis}
\vskip 5pt
\centerline{\it Colloid and Interface Science, Leiden Institute of Chemistry,}
\centerline{\it Gorlaeus Laboratories, P.O. Box 9502, 2300 RA Leiden, The Netherlands.}
\centerline{E-mail: e.blokhuis@chem.leidenuniv.nl}
\vskip 15pt
\centerline{\bf Abstract}
\vskip 5pt
\noindent
Formulas, analogous to the Triezenberg-Zwanzig expression for the surface
tension of a planar interface, are presented for the Tolman length, the
bending rigidity, and the rigidity constant associated with Gaussian curvature.
These expressions feature the Ornstein-Zernike direct correlation function
and are derived from considering the deformation of a liquid drop in the
presence of an external field. This approach is in line with the original
analysis by Yvon in 1948. It is shown that our expressions reduce to previous
results from density functional theory when a mean-field approximation is
made for the direct correlation function. We stress the importance of the
form of the external field used and show how the values of the rigidity
constants depend on it.
\vskip 5pt
\noindent
\centerline{\rule{300pt}{1pt}}

\section{Introduction}
\label{sec-introduction}

\noindent
The most important quantity to define the properties of
a liquid surface is its free energy or {\em surface tension}.
The determination of the surface tension is therefore one of
the key objectives in the Statistical Thermodynamical treatment
of liquid surfaces \cite{RW}. Over the years, different routes have
led to different expressions for the surface tension each
having their own merits and limitations.

At the start of any historical overview are {\em mean-field}
expressions for the surface tension. At the beginning lies the
squared-gradient expression derived by van der Waals in 1893 \cite{vdW}:
\begin{equation}
\label{eq:sigma_vdW}
\sigma = 2 m \int\limits_{-\infty}^{\infty} \!\!\! dz \; \rho_0^{\prime}(z)^2 \,,
\end{equation}
where $\rho_0(z)$ is the density profile of a planar interface. The squared-gradient
expression is a particular result derived from the more general density functional
theory (DFT) \cite{Sullivan, Evans79, Evans90, Evans84}, at the heart of which lies
a division of the interaction potential in a hard sphere reference system perturbed
by attractive forces, $U(r) \!=\! U_{\rm hs}(r) + U_{\rm att}(r)$. For the surface
tension, DFT gives \cite{RW}
\begin{equation}
\label{eq:sigma_DFT}
\sigma = - \frac{1}{4} \int\limits_{-\infty}^{\infty} \!\!\! dz_1 \! \int \!\! d\vec{r}_{12} \;
U_{\rm att}(r) \, r^2 (1-s^2) \, \rho_0^{\prime}(z_1) \rho_0^{\prime}(z_2) \,,
\end{equation}
where the integration over the interparticle distance vector, $\vec{r}_{12} \!\equiv\!
\vec{r}_{2} - \vec{r}_{1}$, is:
\begin{equation}
\int \!\! d\vec{r}_{12} = \int\limits_{0}^{2 \pi} \!\! d\varphi \int\limits_{-1}^{1} \!\! ds
\int\limits_{0}^{\infty} \!\! dr \; r^2 \,,
\end{equation}
which defines $s \!=\! \cos(\theta_{12})$ and $z_2 \!=\! z_1 + r s$. A gradient
expansion of the density in the expression for the surface tension in
Eq.(\ref{eq:sigma_DFT}) yields the van der Waals expression in
Eq.(\ref{eq:sigma_vdW}) with the identification,
$m \!=\! - (1/12) \int \! d\vec{r}_{12} \; r^2 \, U_{\rm att}(r)$.

A second approach results in the so-called {\em virial expression} for the surface
tension, first derived by Kirkwood and Buff in 1948 \cite{KB}:
\begin{equation}
\label{eq:sigma_vir}
\sigma = \frac{1}{4} \int\limits_{-\infty}^{\infty} \!\!\! dz_1 \! \int \!\! d\vec{r}_{12} \;
U^{\prime}(r) \, r (1-3s^2) \, \rho_0^{(2)}(z_1, z_2,r) \,,
\end{equation}
where $\rho_0^{(2)}(z_1, z_2,r)$ is the pair density correlation function of a
planar interface. Even though this expression is derived by Kirkwood and Buff
using a {\em mechanical} approach, by way of the evaluation of the components of the
pressure tensor in the interfacial region \cite{KB}, it is an {\em exact} equation,
the only assumption being made is that of pair-wise additivity of the molecular
interactions. The theoretical evaluation
of the surface tension using the Kirkwood-Buff expression requires
knowledge of the pair correlation function in the interfacial region,
which is hard to access theoretically but can be determined to great
accuracy in computer simulations \cite{Simulations1, Simulations2}. 

A third approach, which also provides an expression for the surface tension
that is exact, results in the so-called Triezenberg-Zwanzig (TZ) expression
\cite{Yvon, TZ}:
\begin{equation}
\label{eq:sigma_TZ}
\sigma = - \frac{k_{\rm B} T}{4} \int\limits_{-\infty}^{\infty} \!\!\! dz_1 \! \int \!\! d\vec{r}_{12} \;
r^2 (1-s^2) \, \rho_0^{\prime}(z_1) \rho_0^{\prime}(z_2) \, C_0(z_1,z_2,r) \,,
\end{equation}
where $C_0(z_1,z_2,r)$ is now the {\em direct correlation function} \cite{OZ}
of the planar interface.

In this article, we are interested in deriving expressions for the
surface tension of a {\em curved surface}. In particular, we investigate
the expansion of the surface tension $\sigma(R)$ to {\em second} order
in the (inverse) radius of curvature $R$ as described by the curvature
coefficients $\delta$ (the Tolman length) \cite{Tolman}, $k$ (the
rigidity constant associated with bending or bending rigidity) and
$\bar{k}$ (the rigidity constant associated with Gaussian curvature).
For a spherical and cylindrical surface, this expansion has the form
\cite{Helfrich}:
\begin{eqnarray}
\label{eq:sigma(R)}
\sigma_s(R) &=& \sigma - \frac{2 \delta \sigma}{R} 
+ \frac{(2 k + \bar{k})}{R^2} + \ldots \hspace*{25pt} {\rm (sphere)} \\
\sigma_c(R) &=& \sigma - \frac{\delta \sigma}{R} 
+ \frac{k}{2 R^2} + \ldots \hspace*{56pt} {\rm (cylinder)} \nonumber
\end{eqnarray}
The virial expressions and mean-field expressions (DFT) for $\delta$,
$k$ and $\bar{k}$ have been derived some 20 years ago \cite{Blokhuis92a,
Blokhuis93, Giessen98}, but until now, the corresponding expressions
in terms of the {\em direct correlation function} are lacking.
Our goal in this article is to fill this void and to derive
Triezenberg-Zwanzig-like expressions for the Tolman length $\delta$,
and the rigidity constants $k$ and $\bar{k}$.
To achieve this, we follow in the footsteps of the pioneering work by
Jim Henderson and coworkers who addressed precisely this problem in three
papers in the early 80's \cite{Hemingway81, Henderson82, Schofield82}.
They derived an expression for the Tolman length in terms of the direct
correlation function, which is consistent with the expression presented here.
Furthermore, they pointed to difficulties in extending the analysis
to {\bf second order} in the expansion in the curvature, i.e. to the
order of the rigidity constants $k$ and $\bar{k}$, which find their
origin in the fact that the thermodynamic conditions used to impose
a certain curvature of the interface then play a role.

The article is organized as follows. In Section \ref{sec-fluct_vs_eq}
we address the Thermodynamics involved in deriving expressions for the rigidity
constants in general. Two distinct thermodynamic routes are formulated which
correspond to the two original derivations of the Triezenberg-Zwanzig
expression for the surface tension of the planar interface:
(i), the {\em fluctuation route}, used by Triezenberg and Zwanzig themselves
in 1972, in which one considers the second order change in free energy
due to surface fluctuations; (ii), the {\em equilibrium route}, used by
Lovett {\em et al.} \cite{Lovett}, following an approach suggested by
Yvon in 1948 \cite{Yvon}, in which one investigates the deformation
of a planar interface induced by an external field.
These two approaches, which result in the same expression for the
surface tension of the planar interface, are discussed in Section
\ref{sec-derivation} and extended to curved surfaces. The full
calculation of the Tolman length $\delta$ and the rigidity constants
$k$ and $\bar{k}$ is presented in Section \ref{sec-expressions},
with details of the calculations left to the Appendices. We end
with a discussion of results.

\section{Fluctuation route versus Equilibrium route}
\label{sec-fluct_vs_eq}

\noindent
Historically, the Triezenberg-Zwanzig expression for the surface tension
has been derived in two distinct ways: (i), by the analysis of the second
order change in free energy due to surface fluctuations \cite{TZ}; (ii),
by applying an external field to deform an initially planar interface
\cite{Yvon, Lovett}. These two routes, which we have termed previously
\cite{Blokhuis99, Blokhuis09} as the {\bf fluctuation route} and the
{\bf equilibrium route}, are connected to the (thermodynamic) path
considered to deform the surface and to induce a certain interfacial
curvature. 

In the {\em fluctuation route}, the curvature of the interface naturally
comes about by the presence of thermal fluctuations. The bulk region
is {\em unaffected} and the chemical potential remains that at two-phase
coexistence, $\mu \!=\! \mu_{\rm coex}$. This route is in many senses
equivalent to a route in which one considers the presence of a 
{\em non-uniform} external field that is non-zero in the interfacial
region only. The ``fluctuation'' is then envisioned as being induced
by the non-uniform external field.

In the {\em equilibrium route}, the interface is curved by changing the
value of the chemical potential to a value {\em off-coexistence} at fixed
temperature ($\mu \!>\! \mu_{\rm coex}$). One then considers the surface
tension $\sigma(R)$ of a spherically (or cylindrically) shaped liquid droplet
with radius $R$ in metastable equilibrium with a bulk vapour. In effect,
a {\em uniform} external field is added of the form $V_{\rm ext}(\vec{r})
\!=\! - \Delta \mu$, where $\Delta \mu \!=\! \mu - \mu_{\rm coex}$.
Since the uniform external field acts throughout the system, it {\em also}
affects the bulk densities far from the interfacial region.

These two distinct routes lead to the {\em same} expression for the
planar surface tension, but it was already noted by Henderson
and coworkers \cite{Hemingway81, Henderson82, Schofield82} that this
may not be the case for the {\em radius dependent} surface tension $\sigma(R)$,
and in particular for the coefficients in an expansion in $1/R$ of
the surface tension {\em beyond} the Tolman length. Their concerns
were connected to the fact that {\em different} external fields may
lead to the {\em same} interfacial curvature and that it is not
a priori clear that the expression for the radius dependent
surface tension derived using one external field equals
that of another. Although the two routes {\em do} lead to the same expression
for the surface tension, it turns out that this indeed is not the case
for the {\em rigidity constants}. This was demonstrated previously
in the context of DFT expressions for the bending rigidity $k$
\cite{Blokhuis09, Blokhuis13}: the fluctuation route was investigated
by supposing the presence of an external field chosen proportional
to the derivative of the density profile to ensure that it acts in
the interfacial region only \cite{Blokhuis09}. It was shown that the
value obtained for the bending rigidity then {\em differs} in magnitude
and scaling behaviour from that obtained using the equilibrium route
\cite{Blokhuis13} (see Appendix \ref{app-DFT}).

That the thermodynamic conditions imposed to induce a certain
interfacial curvature have bearing on the rigidity constants,
may also be deduced from the fact that the rigidity constants
depend on the density profile $\rho_1(z)$, which describes the
leading order change in the distribution of molecules when the
interface is curved. The fact that the density profiles $\rho_1(z)$
are {\em different} in the fluctuation and equilibrium route,
follows directly from the observation that in the equilibrium route,
$\rho_1(z)$ is not equal to zero in the bulk regions since the
chemical potential is also different ($\mu \!\neq\! \mu_{\rm coex}$),
while it {\em is} zero in the fluctuation route since fluctuations
of the interface do not affect the thermodynamic state of the bulk
regions.

With this important observation in mind, we consider, in the
next section the usual derivations of the Triezenberg-Zwanzig
expression for the surface tension of a planar interface and
discuss how they are extended to determine the Tolman length
$\delta$ and the rigidity constants $k$ and $\bar{k}$.

\section{Two derivations of the Triezenberg-Zwanzig expression for the surface tension}
\label{sec-derivation}

\noindent
To understand how the direct correlation function enters in the
Triezenberg-Zwanzig expression for the surface tension, we need
to address the Statistical Thermodynamical definition of the direct
correlation function \cite{OZ}.

In the fluctuation route considered first, the direct correlation function
describes the (second order) response of the free energy due to density fluctuations.
This route was used in the original derivation by Triezenberg and
Zwanzig \cite{TZ} in 1972. In the {\em equilibrium} route, the direct
correlation function describes how a small change in the local density
can be interpreted as a change in the local external field (second
Yvon equation). This approach, inspired by the work of Yvon in 1948 \cite{Yvon},
was used by Lovett {\em et al.} \cite{Lovett} to rederive the TZ expression
for the surface tension of the planar interface by considering the external
field necessary to bend an initially flat surface into a spherical droplet.
This procedure was then extended by Henderson and Schofield \cite{Henderson82}
to spherical droplets in order to have access to the full radius dependent
surface tension.

\subsection{Fluctuation route}

\noindent
In the fluctuation route, one investigates the change in free energy
due to density fluctuations around a {\em planar} interface. The direct
correlation function describes the second order change in the free
energy in response to such a density fluctuation:
\begin{equation}
\label{eq:Delta_Omega_TZ}
\Delta \Omega = \frac{k_{\rm B} T}{2} \int \!\! d\vec{r}_1 \! \int \!\! d\vec{r}_{12} \;
C_0(z_1, z_2,r) \, \delta \rho(\vec{r}_1) \, \delta \rho(\vec{r}_2) \,.
\end{equation} 
The density fluctuation can be expressed in terms of a {\em height function}
$h(\vec{r}_{\parallel})$ that describes the shift in the location of the
surface \cite{Parry, Blokhuis09}:
\begin{equation}
\label{eq:delta_rho}
\delta \rho(\vec{r}) \equiv \rho(\vec{r}) - \rho_0(z) =
- \rho^{\prime}_0(z) \, h(\vec{r}_{\parallel})
- \frac{\rho_1(z)}{2} \, \vec{\nabla}^2 h(\vec{r}_{\parallel}) + \ldots \,,
\end{equation}
where it is assumed that $\mid\! \vec{\nabla} h \!\mid \ll 1$
and where $\rho_1(z)$ is identified as the correction to the
density profile due to the {\em curvature} of the interface. 
In general, $\rho_1(z)$ describes the rearrangement of molecules
that results when the interface is curved, but its form still depends
on the {\em way} the curvature is brought upon.

The surface tension and bending rigidity are thermodynamically defined
as the change in free energy due to height fluctuations \cite{Meunier}:
\begin{equation}
\label{eq:Omega_ECW}
\Delta \Omega = \frac{1}{2} \int \!\! \frac{d\vec{q}}{(2 \pi)^2} \;
[ \; \sigma \, q^2 + k \, q^4 + \ldots \; ] \, h(\vec{q}) \, h(-\vec{q}) \,.
\end{equation}
Inserting Eq.(\ref{eq:delta_rho}) into Eq.(\ref{eq:Delta_Omega_TZ})
and comparing the result to Eq.(\ref{eq:Omega_ECW}), we retrieve the
TZ expression for the surface tension in Eq.(\ref{eq:sigma_TZ}).
A further result is that for the bending rigidity one finds \cite{Parry}:
\begin{eqnarray}
\label{eq:k_TZ}
k &=& \frac{k_{\rm B} T}{4} \int\limits_{-\infty}^{\infty} \!\!\! dz_1 \! \int \!\! d\vec{r}_{12} \;
\rho_1(z_1) \rho_1(z_2) \, C_0(z_1, z_2, r) \\
&& + \frac{k_{\rm B} T}{4} \int\limits_{-\infty}^{\infty} \!\!\! dz_1 \! \int \!\! d\vec{r}_{12} \;
r^2 (1-s^2) \, \rho_0^{\prime}(z_1) \rho_1(z_2) \, C_0(z_1, z_2, r) \nonumber \\
&& + \frac{k_{\rm B} T}{64} \int\limits_{-\infty}^{\infty} \!\!\! dz_1 \! \int \!\! d\vec{r}_{12} \;
r^4 (1-s^2)^2 \, \rho_0^{\prime}(z_1) \rho_0^{\prime}(z_2) \, C_0(z_1, z_2, r) \,. \nonumber
\end{eqnarray}
This formal expression for $k$ was first derived by Parry and Boulter \cite{Parry}.
Even though this expression is exact, for the explicit evaluation of it,
one still requires some way of determining the density profile $\rho_1(z)$. 
Furthermore, expressions for $\delta$ and $\bar{k}$ {\em cannot} be derived
using this route.

\subsection{Equilibrium route}

\noindent
To understand the way the direct correlation function enters the expression
for the surface tension in the equilibrium route, we need to discuss the
two Yvon equations which are in the heart of any Statistical Thermodynamics
treatment of surfaces \cite{RW, Evans79, Evans90}. In the first Yvon equation,
the {\em total} correlation function, $G(\vec{r}_1, \vec{r}_2)$, relates a
small change in the local external field to a change in the local density
\cite{Evans79, Evans90}:
\begin{equation}
\label{eq:Yvon_1}
\delta \rho (\vec{r}_1) = - \frac{1}{k_{\rm B} T} \int \!\! d\vec{r}_2 \;
G(\vec{r}_1,\vec{r}_2) \, \delta V_{\rm ext}(\vec{r}_2) \,,
\end{equation}
where $G(\vec{r}_1,\vec{r}_2)$ is related to the pair density as:
\begin{equation}
\label{eq:G}
G(\vec{r}_1,\vec{r}_2) = \rho^{(2)}(\vec{r}_1,\vec{r}_2)
- \rho(\vec{r}_1) \, \rho(\vec{r}_2) + \rho(\vec{r}_1) \, \delta(\vec{r}_2 - \vec{r}_1) \,.
\end{equation}
The inverse of the total correlation function defines the {\em direct
correlation function} \cite{OZ} and it describes in the {\em second Yvon
equation} how a small change in the local density can be attributed to
a change in the local external field \cite{Evans79, Evans90}:
\begin{equation}
\label{eq:Yvon_2}
\delta V_{\rm ext}(\vec{r}_1) = - k_{\rm B} T \int \!\! d\vec{r}_2 \;
C(\vec{r}_1,\vec{r}_2) \, \delta \rho(\vec{r}_2) \,.
\end{equation}
By considering translational deformations, the second Yvon equation
appears in the following form:
\begin{equation}
\label{eq:Yvon_2_trans}
\vec{\nabla}_1 V_{\rm ext}(\vec{r}_1) = - k_{\rm B} T \int \!\! d\vec{r}_2 \;
C(\vec{r}_1,\vec{r}_2) \, \vec{\nabla}_2 \, \rho(\vec{r}_2) \,,
\end{equation}
which, in planar geometry, reduces to:
\begin{equation}
\label{eq:Yvon_2_p}
V_{\rm ext}^{\prime}(z_1) = - k_{\rm B} T \int \!\! d\vec{r}_2 \;
C_0(z_1,z_2,r) \, \rho_0^{\prime}(z_2) \,.
\end{equation}

As a next step, we consider the influence of the external field on the
{\em Laplace pressure difference} $\Delta p$ across a spherical or cylindrical
droplet with radius $R$. The generalized Laplace equation for the pressure
difference across a surface with (constant) curvatures $J \!=\! 1/R_1 + 1/R_2$
and $K \!=\! 1/(R_1 R_2)$ is given by \cite{Shape_equation}:
\begin{equation}
\label{eq:Laplace}
\Delta p = \sigma \, J - 2 \delta \sigma \, K - \frac{k}{2} \, J^3 + 2 k \, J \, K + \ldots
\end{equation}
For a spherically or cylindrically shaped surface, this expansion takes the form:
\begin{eqnarray}
\label{eq:Delta_p_s}
\Delta p &=& \frac{2 \sigma}{R} - \frac{2 \delta \sigma}{R^2} + \dots \hspace*{40pt} {\rm (sphere)} \\
\label{eq:Delta_p_c}
\Delta p &=& \frac{\sigma}{R} - \frac{k}{2 R^3} + \dots \hspace*{43pt} {\rm (cylinder)}
\end{eqnarray}
where the dots represent terms of order ${\cal O}(1/R^4)$, which indicates that
the term proportional to $1/R^3$ is absent in the expansion of the spherical
interface. It is important to mention that these expressions are derived with
the {\em equimolar} radius \cite{Gibbs} chosen as the radius of the spherically
(or cylindrically) shaped surface, i.e. $R \!=\! R_e$, where $R_e$ for a
spherical liquid droplet is defined by:
\begin{equation}
\label{eq:R}
4 \pi \int\limits_{0}^{\infty} \!\! dr \; r^2 \left[ \, \rho_s(r) - \rho_v \right]
= \frac{4 \pi}{3} \, R_e^3 \, (\rho_{\ell} - \rho_v) \,.
\end{equation}

Next, we consider the consequences of {\em mechanical balance} in the presence
of an external field. The condition of mechanical balance can be expressed
in terms of the pressure tensor \cite{RW}:
\begin{equation}
\label{eq:pressure_balance}
\vec{\nabla} \cdot \vec{\vec{\rm p}} (\vec{r}) =
- \rho(\vec{r}) \, \vec{\nabla} V_{\rm ext}(\vec{r}) \,.
\end{equation}
In {\em spherical symmetry}, the expression for mechanical balance leads to:
\begin{equation}
\label{eq:pressure_components}
p_{\rm N}^{\prime}(r) = \frac{2}{r} [ \; p_{\rm T}(r) - p_{\rm N}(r) \; ]
-  \rho_s(r) \, V^{\prime}_{\rm ext}(r) \,,
\end{equation}
which integrated from inside the liquid to the vapour phase gives
\begin{equation}
\label{eq:integrated_pressure}
\Delta p = \int\limits_{0}^{\infty} \!\! dr \; \frac{2}{r} [ \; p_{\rm N}(r) - p_{\rm T}(r) \; ]
+ \int\limits_{0}^{\infty} \!\! dr \; \rho_s(r) \, V^{\prime}_{\rm ext}(r) \,.
\end{equation}
Following the derivation by Yvon \cite{Yvon}, this expression was applied
by Lovett {\em et al.} \cite{Lovett} to rederive the TZ expression for the
surface tension by considering the situation in which an initially flat surface,
in the absence of an external field, is bend into a spherical droplet by an
infinitesimally small external field $\delta V_{\rm ext}(r)$ in such a way
that {\em the bulk regions are unaffected}. More generally, one may consider
an already curved spherical droplet, whose curvature is increased or
decreased ($R \!\rightarrow\! R + \delta R$) by the infinitesimally small
external field \cite{Henderson82}. In either case, the change in pressure
difference is zero, $\delta (\Delta p) \!=\! 0$, and the two terms on
the right-hand-side of Eq.(\ref{eq:integrated_pressure}) must balance.
Since we have that in the {\em absence} of the external field
\begin{equation}
\label{eq:Laplace_s_1}
\Delta p = \int\limits_{0}^{\infty} \!\! dr \; \frac{2}{r} [ \; p_{\rm N}(r) - p_{\rm T}(r) \; ]
= \frac{2 \sigma}{R} - \frac{2 \delta \sigma}{R^2} + \dots \,,
\end{equation}
this implies that
\begin{equation}
\label{eq:Laplace_s_2}
\delta \left( \frac{2 \sigma}{R} - \frac{2 \delta \sigma}{R^2} + \dots \right)
= \int\limits_{0}^{\infty} \!\! dr \; \rho_s^{\prime}(r) \, \delta V_{\rm ext}(r) \,,
\end{equation}
where we have integrated the last term in Eq.(\ref{eq:integrated_pressure})
by parts. The second Yvon equation in Eq.(\ref{eq:Yvon_2}) is then used to
rewrite the right-hand-side as:
\begin{equation}
\label{eq:Laplace_s_3}
\left( \frac{2 \sigma}{R^2} - \frac{4 \delta \sigma}{R^3} + \ldots \right) \delta R
= k_{\rm B} T \int\limits_{0}^{\infty} \!\! dr_1 \int \!\! d\vec{r}_2 \;
C_s(r_1,r_2,r) \, \rho_s^{\prime}(r_1) \, \delta \rho_s(r_2) \,.
\end{equation}
The leading order term in the expansion in $1/R$ of the right-hand-side in
this equation leads to the TZ expression for $\sigma$ \cite{Lovett}, while
the next to leading order term leads to an expression for the Tolman length
\cite{Hemingway81, Henderson82}. Here, we are interested in the {\em second
order} terms beyond the leading term. Even though the coefficient of the
$1/R^4$ term in Eq.(\ref{eq:Laplace_s_3}) happens to be zero, its absence
shall provide information for the determination of the combination
of rigidity constants $2 k + \bar{k}$.

In order to obtain an expression for the bending rigidity $k$ separately, the
analysis is also carried out for a {\em cylindrically} shaped interface.
In cylindrical geometry, the consideration of mechanical balance than
ultimately leads to
\begin{equation}
\label{eq:Laplace_c_1}
\left( \frac{\sigma}{R^2} - \frac{3 k}{2 R^4} + \ldots \right) \delta R
= k_{\rm B} T \int\limits_{0}^{\infty} \!\! dr_1 \int \!\! d\vec{r}_2 \;
C_c(r_1,r_2,\varphi,r) \, \rho_c^{\prime}(r_1) \, \delta \rho_c(r_2) \,.
\end{equation}
For the cylindrical interface, the $1/R^4$ term is related to the
bending rigidity $k$ and it is the $1/R^3$ term that vanishes. In the
next section, Eqs.(\ref{eq:Laplace_s_3}) and (\ref{eq:Laplace_c_1})
are used to carry out a systematic expansion in $1/R$ to derive explicit
expressions for the Tolman length $\delta$, and the rigidity constants
$k$ and $\bar{k}$.

\section{Equilibrium route to determine $\delta$, $k$, and $\bar{k}$}
\label{sec-expressions}

\noindent
We consider the situation outlined in the previous section: a spherical
liquid droplet with radius $R$ is deformed by an infinitesimally
small external field into another spherical droplet with radius
$R \!\rightarrow\! R + \delta R$ without affecting the bulk regions.
To determine the resulting change in density profile, we use the fact
that we can write the density as $\rho_s(r) \!=\! \rho_s(r - R ; R)$.
This means that when $R$ is changed, the density profile is affected in
two ways: (i), the variable $z \!\equiv\! r - R$ is shifted and (ii), the
density profile itself depends on $R$. This means that
\begin{eqnarray}
\label{eq:delta_rho_0}
\delta \rho_s(r) &=& \rho_s(r - R - \delta R ; R + \delta R) - \rho_s(r - R ; R) \nonumber \\
&=& - \rho_s^{\prime}(r) \, \delta R + \rho_s(r; R + \delta R) - \rho_s(r; R) \,.
\end{eqnarray}
When we expand $\rho_s(r)$ in $1/R$, as in Eq.(\ref{eq:expansion_rho}),
this gives
\begin{equation}
\label{eq:delta_rho_1}
\delta \rho_s(r) = - \left[ \, \rho_s^{\prime}(r)
+ \frac{\rho_1(r)}{R^2} + \frac{2 \, \rho_{s,2}(r)}{R^3} + \ldots \right] \delta R \,.
\end{equation}
This expression is inserted into Eq.(\ref{eq:Laplace_s_3}) and expanded
to second order beyond the leading order term. Details of the analysis for
spherical droplets are outlined in Appendix \ref{app-sphere}; details of the
analysis for cylindrical droplets in Appendix \ref{app-cylinder}. For the
Tolman length, the analysis of the spherical interface in Appendix \ref{app-sphere}
leads to Eq.(\ref{eq:delta_s}):
\begin{equation}
\label{eq:delta_1}
\delta \sigma = \frac{k_{\rm B} T}{8} \int\limits_{-\infty}^{\infty} \!\!\! dz_1 \! \int \!\! d\vec{r}_{12} \;
r^2 (1-s^2) \, \left[ \, \rho_s^{\prime}(z_1) \rho_s^{\prime}(z_2) \, C_s(z_1,z_2,r) \, \right]_1 \,,
\end{equation}
where the subscript $1$ refers to the leading order correction in an expansion
in $1/R$ (see Eq.(\ref{eq:expansion_rho})). The evaluation of the Tolman length
using this expression therefore requires knowledge of the way the density profile
and direct correlation function depend on the radius of curvature.

The expression for the Tolman length in Eq.(\ref{eq:delta_1}) can also be derived
from the expression for $\sigma_s(R)$ presented by Henderson and coworkers in
refs.~\cite{Hemingway81,Henderson82}:
\begin{equation}
\label{eq:sigma_Henderson}
\sigma_s(R) = - \frac{k_{\rm B} T}{4} \int\limits_{0}^{\infty} \!\! dr_1 \! \int \! d\vec{r}_{12} \;
\left[ r^2 - (r_1 - r_2)^2 \right] \, \rho_s^{\prime}(r_1) \rho_s^{\prime}(r_2) \, C_s(r_1,r_2,r) \,,
\end{equation}
which is constructed to be correct to order ${\cal O}(1/R)$ \cite{Henderson82}.
An explicit expansion in $1/R$ of the term in square brackets gives:
\begin{equation}
r^2 - (r_1 - r_2)^2 = r^2 (1-s^2) \, (1 - \frac{sr}{R}) + \ldots \,.
\end{equation}
Inserting this expansion in Eq.(\ref{eq:sigma_Henderson}) and comparing
with the expansion for $\sigma_s(R)$ in Eq.(\ref{eq:sigma(R)}) results in
the expression for the Tolman length in Eq.(\ref{eq:delta_1}). It should be
mentioned that starting with the Henderson expression in Eq.(\ref{eq:sigma_Henderson}),
an expression for the Tolman length very similar to Eq.(\ref{eq:delta_1}) was derived
by Barrett in 1999 \cite{Barrett99}. The only difference is that in the Barrett
expression for the Tolman length, $C_{s,1}(z_1,z_2,r)$ in Eq.(\ref{eq:delta_1})
is rewritten in terms of the {\em triplet} direct correlation function of
the planar interface \cite{Barrett99}.

When the expression for the Tolman length in Eq.(\ref{eq:delta_1}) is combined
with the result of the analysis of the {\em cylindrical} interface in Appendix
\ref{app-cylinder}, we find as an alternative expression for the Tolman length:
\begin{equation}
\label{eq:delta_2}
\delta \sigma = \frac{k_{\rm B} T}{4} \int\limits_{-\infty}^{\infty} \!\!\! dz_1 \! \int \!\! d\vec{r}_{12} \;
r^2 (1-s^2) \, z_1 \, \rho_0^{\prime}(z_1) \rho_0^{\prime}(z_2) \, C_0(z_1,z_2,r) \,.
\end{equation}
It is reminded that in this expression the location of the $z \!=\! 0$ plane
corresponds to the {\em equimolar} surface. The advantage of this expression for
the Tolman length is that it can be evaluated from the properties of the {\em planar}
interface only. In Appendix \ref{app-DFT} we verify that both expressions
for the Tolman length in Eqs.(\ref{eq:delta_1}) and (\ref{eq:delta_2})
are consistent with known DFT expressions \cite{Giessen98, Blokhuis13}.

The vanishing of the $1/R^4$ term in Eq.(\ref{eq:Laplace_s_3}) in the analysis for
the spherical interface, leads to Eq.(\ref{eq:2k_kbar_s}) in Appendix \ref{app-sphere}. 
This gives the following two expressions for the combination $2 k + \bar{k}$:
\begin{eqnarray}
\label{eq:2k_kbar_1}
2 k + \bar{k} &=& - \frac{k_{\rm B} T}{12} \int\limits_{-\infty}^{\infty} \!\!\! dz_1 \! \int \!\! d\vec{r}_{12} \;
r^2 (1-s^2) \, \left[ \, \rho_s^{\prime}(z_1) \rho_s^{\prime}(z_2) \, C_s(z_1,z_2,r) \, \right]_2 \\
&& - \frac{2 \, k_{\rm B} T}{3} \int\limits_{-\infty}^{\infty} \!\!\! dz_1 \! \int \!\! d\vec{r}_{12} \;
rs \, \rho_0^{\prime}(z_1) \rho_{s,2}(z_2) \, C_0(z_1,z_2,r) \nonumber \\
&& - \frac{k_{\rm B} T}{3} \int\limits_{-\infty}^{\infty} \!\!\! dz_1 \! \int \!\! d\vec{r}_{12} \;
rs \, \rho_1(z_2) \, \left[ \, \rho_s^{\prime}(z_1) \, C_s(z_1,z_2,r) \, \right]_1 \nonumber \\
&& - \frac{k_{\rm B} T}{4} \int\limits_{-\infty}^{\infty} \!\!\! dz_1 \! \int \!\! d\vec{r}_{12} \;
r^2 (1-s^2) \, z_1^2 \, \rho_0^{\prime}(z_1) \rho_0^{\prime}(z_2) \, C_0(z_1,z_2,r) \nonumber \\
&& + \frac{k_{\rm B} T}{48} \int\limits_{-\infty}^{\infty} \!\!\! dz_1 \! \int \!\! d\vec{r}_{12} \;
r^4 (1-s^4) \, \rho_0^{\prime}(z_1) \rho_0^{\prime}(z_2) \, C_0(z_1,z_2,r) \,, \nonumber
\end{eqnarray}
and
\begin{eqnarray}
\label{eq:2k_kbar_2}
2 k + \bar{k} &=& - \frac{k_{\rm B} T}{6} \int\limits_{-\infty}^{\infty} \!\!\! dz_1 \! \int \!\! d\vec{r}_{12} \;
r^2 (1-s^2) \, z_1 \, \left[ \, \rho_s^{\prime}(z_1) \rho_s^{\prime}(z_2) \, C_s(z_1,z_2,r) \, \right]_1 \\
&& - \frac{k_{\rm B} T}{3} \int\limits_{-\infty}^{\infty} \!\!\! dz_1 \! \int \!\! d\vec{r}_{12} \;
rs \, z_1 \, \rho_0^{\prime}(z_1) \rho_1(z_2) \, C_0(z_1,z_2,r) \nonumber \\
&& + \frac{k_{\rm B} T}{12} \int\limits_{-\infty}^{\infty} \!\!\! dz_1 \! \int \!\! d\vec{r}_{12} \;
r^2 (1-3s^2) \, \rho_0^{\prime}(z_1) \rho_1(z_2) \, C_0(z_1,z_2,r) \nonumber \\
&& + \frac{k_{\rm B} T}{48} \int\limits_{-\infty}^{\infty} \!\!\! dz_1 \! \int \!\! d\vec{r}_{12} \;
r^4 (1-s^4) \, \rho_0^{\prime}(z_1) \rho_0^{\prime}(z_2) \, C_0(z_1,z_2,r) \,. \nonumber
\end{eqnarray}

The analysis for the $1/R^4$ term in Eq.(\ref{eq:Laplace_c_1}) for the
cylindrical interface, leads to  Eq.(\ref{eq:k_c}) in Appendix \ref{app-cylinder}.
This gives the following two expressions for the bending rigidity $k$:
\begin{eqnarray}
\label{eq:k_1}
k &=& - \frac{k_{\rm B} T}{3} \int\limits_{-\infty}^{\infty} \!\!\! dz_1 \! \int \!\! d\vec{r}_{12} \;
r^2 (1-s^2) \, \sin^2\!\varphi \, \left[ \, \rho_c^{\prime}(z_1) \rho_c^{\prime}(z_2)
\, C_c(z_1,z_2,\varphi,r) \, \right]_2 \\
&& - \frac{4 \, k_{\rm B} T}{3} \int\limits_{-\infty}^{\infty} \!\!\! dz_1 \! \int \!\! d\vec{r}_{12} \;
rs \, \rho_0^{\prime}(z_1) \rho_{c,2}(z_2) \, C_0(z_1,z_2,r) \nonumber \\
&& - \frac{k_{\rm B} T}{6} \int\limits_{-\infty}^{\infty} \!\!\! dz_1 \! \int \!\! d\vec{r}_{12} \;
rs \, \rho_1(z_2) \, \left[ \, \rho_s^{\prime}(z_1) \, C_s(z_1,z_2,r) \, \right]_1 \nonumber \\
&& - \frac{k_{\rm B} T}{4} \int\limits_{-\infty}^{\infty} \!\!\! dz_1 \! \int \!\! d\vec{r}_{12} \;
r^2 (1-s^2) \, z_1^2 \, \rho_0^{\prime}(z_1) \rho_0^{\prime}(z_2) \, C_0(z_1,z_2,r) \nonumber \\
&& + \frac{k_{\rm B} T}{64} \int\limits_{-\infty}^{\infty} \!\!\! dz_1 \! \int \!\! d\vec{r}_{12} \;
r^4 (1-s^2) (1+3s^2) \, \rho_0^{\prime}(z_1) \rho_0^{\prime}(z_2) \, C_0(z_1,z_2,r) \,, \nonumber
\end{eqnarray}
and
\begin{eqnarray}
\label{eq:k_2}
k &=& - \frac{k_{\rm B} T}{12} \int\limits_{-\infty}^{\infty} \!\!\! dz_1 \! \int \!\! d\vec{r}_{12} \;
r^2 (1-s^2) \, z_1 \left[ \, \rho_s^{\prime}(z_1) \rho_s^{\prime}(z_2) \, C_s(z_1,z_2,r) \, \right]_1 \\
&& - \frac{k_{\rm B} T}{6} \int\limits_{-\infty}^{\infty} \!\!\! dz_1 \! \int \!\! d\vec{r}_{12} \;
rs \, z_1 \, \rho_0^{\prime}(z_1) \rho_1(z_2) \, C_0(z_1,z_2,r) \nonumber \\
&& + \frac{k_{\rm B} T}{24} \int\limits_{-\infty}^{\infty} \!\!\! dz_1 \! \int \!\! d\vec{r}_{12} \;
r^2 (1-3s^2) \, \rho_0^{\prime}(z_1) \rho_1(z_2) \, C_0(z_1,z_2,r) \nonumber \\
&& + \frac{k_{\rm B} T}{8} \int\limits_{-\infty}^{\infty} \!\!\! dz_1 \! \int \!\! d\vec{r}_{12} \;
r^2 (1-s^2) \, z_1^2 \, \rho_0^{\prime}(z_1) \rho_0^{\prime}(z_2) \, C_0(z_1,z_2,r) \nonumber \\
&& + \frac{k_{\rm B} T}{64} \int\limits_{-\infty}^{\infty} \!\!\! dz_1 \! \int \!\! d\vec{r}_{12} \;
r^4 (1-s^2)^2 \, \rho_0^{\prime}(z_1) \rho_0^{\prime}(z_2) \, C_0(z_1,z_2,r) \,. \nonumber
\end{eqnarray}
Combining Eqs.(\ref{eq:2k_kbar_2}) and (\ref{eq:k_2}) then leads to the following
expression for $\bar{k}$:
\begin{eqnarray}
\label{eq:k_bar}
\bar{k} &=& - \frac{k_{\rm B} T}{4} \int\limits_{-\infty}^{\infty} \!\!\! dz_1 \! \int \!\! d\vec{r}_{12} \;
r^2 (1-s^2) \, z_1^2 \, \rho_0^{\prime}(z_1) \rho_0^{\prime}(z_2) \, C_0(z_1,z_2,r) \\
&& - \frac{k_{\rm B} T}{96} \int\limits_{-\infty}^{\infty} \!\!\! dz_1 \! \int \!\! d\vec{r}_{12} \;
r^4 (1-s^2) (1-5s^2) \, \rho_0^{\prime}(z_1) \rho_0^{\prime}(z_2) \, C_0(z_1,z_2,r) \,. \nonumber
\end{eqnarray}
The expression for the Tolman length in Eq.(\ref{eq:delta_2}) and the expressions
for the rigidity constants in Eqs.(\ref{eq:k_2}) and (\ref{eq:k_bar}) are the final
results of this article.

\section{Discussion}

\noindent
For the surface tension of the planar interface, one may distinguish three
type of microscopic expressions:

\begin{itemize}

\item{Mean-field or DFT expressions of
which the van der Waals squared-gradient expression is an example. These are
approximate in nature but are widely used since they are relatively easy
to evaluate (numerically).}

\item{The Kirkwood-Buff virial expression, which is exact and widely used to
determine the surface tension in computer simulations.}

\item{The Triezenberg-Zwanzig expression in terms of the direct correlation
function. Although the evaluation of the surface tension using this
expression requires access to the direct correlation function, it has
the advantage that it is an {\em exact} expression and the short-ranged
nature of the direct correlation function has proven to be helpful to
settle fundamental issues such as the influence of capillary waves on
the structure and tension of a planar interface in the limit of vanishing
gravitational field \cite{Weeks1, Weeks2, Weeks3}.}

\end{itemize}

\noindent
There exists a great level of consistency between these three type of
expressions for the surface tension. Although it is quite elaborate to
show the equivalence of the Triezenberg-Zwanzig and Kirkwood-Buff expression
for the surface tension \cite{TZ-KB}, it is quite more simple to show
how they are both consistent with mean-field expressions.

Ever since the introduction of the coefficients $\delta$, $k$ and $\bar{k}$ 
that appear in the curvature expansion of the surface tension by Tolman
\cite{Tolman} and then later by Helfrich \cite{Helfrich}, it
has been the goal to formulate the same type of microscopic expressions
and to achieve the same level of consistency as it exists for the surface
tension of the planar interface. 
The road to the realization of this goal started with the formulation
of {\em squared-gradient} expressions, first for the Tolman length by 
Fisher and Wortis in 1984 \cite{FW} and then for the rigidity constants
by Gompper and Zschocke \cite{Gompper} and by us \cite{Blokhuis93} in
the early 90's. These squared-gradient expressions were then
shown to be embedded in more general DFT expressions in 1998
\cite{Giessen98, Blokhuis13}. By that time, also the {\em virial
expressions} for $\delta$, $k$ and $\bar{k}$, analogous to the
Kirkwood-Buff formula for the surface tension, were formulated \cite{Blokhuis92a}
and it was verified that they reduce to the DFT expressions
when a mean-field approximation for the pair density is made.

Important progress was also made in the formulation of Triezenberg-Zwanzig-like
expressions in terms of the {\em direct correlation function}.
Even before the work by Fisher and Wortis in 1984, Henderson and
coworkers \cite{Hemingway81, Henderson82, Schofield82} derived
an expression for $\sigma(R)$ correct to first order in the expansion
in $1/R$. Although it was not shown explicitly at the time, the expression
for the Tolman length in Eq.(\ref{eq:delta_1}) can be extracted from their
analysis, even though the reduction to the more simple expression in
Eq.(\ref{eq:delta_2}) could not yet be made. Further progress was then made
by Parry and Boulter \cite{Parry}, who extended the original analysis
by Triezenberg-Zwanzig and derived the analogous expression for the
bending rigidity in Eq.(\ref{eq:k_TZ}). To understand the status of
this expression, it is important to realize that the very nature of
the rigidity constants $k$ and $\bar{k}$ as the change in free energy
due to the curvature of the interface, immediately leads to the
following two rather subtle issues:

\begin{itemize}

\item{The curvature of the surface, as described by the total and Gaussian
curvatures $J$ and $K$, is {\em not uniquely} defined simply because the
allocation of the position of the dividing surface cannot be made unambiguously.
Always, some procedure to locate the dividing surface either by an
integral constraint (equimolar surface) or by a crossing constraint
has to be chosen. For the surface tension (and Tolman length),
this observation bears no consequences but the rigidity constants
are intrinsically linked to some choice for the location of the dividing
surface. For example, the expressions for $k$ and $\bar{k}$ in Eqs.(\ref{eq:k_2})
and (\ref{eq:k_bar}) are derived using the equimolar surface as dividing
surface (Eq.(\ref{eq:equimolar})).}

\item{The system's response to curvature, and therefore the value of
the rigidity constants, depends on the way this curvature is induced.
Most straightforwardly it is induced by uniformly changing the value of the
chemical potential ({\em equilibrium} route) -- the mean-field and virial
expressions are all derived in this way -- but one could also envision
the application of a non-uniform external field ({\em fluctuation} route).
The corresponding values of the rigidity constants then {\em differ} as
demonstrated previously in the context of DFT \cite{Blokhuis99, Blokhuis06}.
The implication is that the equivalence of the equilibrium route and
the fluctuation route, as it exists for the derivation of the Triezenberg-Zwanzig
expression for the surface tension of the planar interface, does {\em not}
exist for the derivation of Triezenberg-Zwanzig-like expression
for the rigidity constants.}

\end{itemize}

\noindent
The expression for the bending rigidity by Parry and Boulter \cite{Parry}
is derived in the fluctuation route, which leaves the derivation of
Triezenberg-Zwanzig-like expressions for the Tolman length $\delta$
and the rigidity constants $k$ and $\bar{k}$  in the {\em equilibrium}
route as the final piece of the puzzle. In this article, the expressions
sought after are finally derived by extending the analysis by Henderson
and coworkers \cite{Hemingway81, Henderson82, Schofield82}. For the Tolman
length, our final result is the expression in Eq.(\ref{eq:delta_2}).
For the rigidity constants our final expressions are presented in
Eqs.(\ref{eq:k_2}) and (\ref{eq:k_bar}).

The Triezenberg-Zwanzig-like expressions for the Tolman length
and rigidity constants possess, like the Triezenberg-Zwanzig expression
for the surface tension of the planar interface itself, certain
advantages with respect to other approaches. Foremost, the
TZ-like expressions are {\em exact} expressions which means
that they are valid beyond the mean-field approximation.
This is not merely a quantitative issue since it is well-known
that the presence of capillary waves are not fully incorporated
in mean-field theory as it fails, for instance, to capture
the divergence of the interfacial width in the limit
of vanishing gravitational field \cite{Evans79, Weeks1}.
A further advantage of the TZ-like expressions is that the
direct correlation function featured is {\em short-ranged}
even in the situation that the capillary length diverges and
the range of the pair density correlation function becomes
infinite. This is instrumental in addressing the influence of
capillary waves on the surface tension of non-planar interfaces.
For the surface tension of the planar interface it is well-established
that capillary waves reduce the mean-field value for the surface tension
by as much as 20\%, but the same analysis for the Tolman length
or the rigidity constants has not been carried out thus far.
\vskip 15pt
\noindent
{\Large\bf Acknowledgment}
\vskip 5pt
\noindent
This article is dedicated to the occasion of the retirement of Jim Henderson,
whose scientific heritage has been and still is a source of insight and
inspiration to me.

%%%%%%%%%%%%%%%%%%%%%%%%%%%%%%%%%%%%%%%%%%%%%%%%%%%%%%%%%%%%%%%%%%%%%%%%%%%%%%%%%%%%%%%%%%%%%%%%%%%%%%%%%%%%%%

\appendix

\section{Density functional theory expressions}
\label{app-DFT}

\noindent
In this appendix, we present the results of density functional theory
for $\delta$, $k$ and $\bar{k}$ (details of the analysis can be found
in ref.~\cite{Blokhuis13}) and show how the Triezenberg-Zwanzig-like
expressions in terms of the direct correlation function reduce to the
DFT expressions when a mean-field approximation is made. 

The expression for the (grand) free energy in DFT is based on
the division into a hard-sphere reference system plus attractive
forces described by an interaction potential $U_{\rm att}(r)$.
It is the following functional of the density \cite{Sullivan,
Evans79, Evans90, Evans84}:
\begin{equation}
\label{eq:Omega_DFT}
\Omega[\rho] = \int \!\! d\vec{r} \; [ \; f_{\rm hs}(\rho) - \mu \rho(\vec{r}) \; ]
+ \frac{1}{2} \int \!\! d\vec{r}_1 \! \int \!\! d\vec{r}_{12} \;
U_{\rm att}(r) \, \rho(\vec{r}_1) \rho(\vec{r}_2) \,,
\end{equation} 
where $f_{\rm hs}(\rho)$ is the free energy density of the hard-sphere
reference system with uniform density $\rho$. This functional is based
on a local density approximation for the hard-sphere reference fluid,
but more sophisticated approaches exist using weighted-density
\cite{Evans84} or fundamental measure theory \cite{FMT}.

The Euler-Lagrange equation that minimizes the above free energy is given by:
\begin{equation}
\label{eq:EL_DFT}
f^{\prime}_{\rm hs}(\rho) = \mu - \int \!\! d\vec{r}_{12} \; U_{\rm att}(r) \, \rho(\vec{r}_2) \,.
\end{equation}
To derive expressions for the curvature coefficients $\delta$, $k$
and $\bar{k}$, one needs to consider the expansion of the
density and free energy in Eq.(\ref{eq:Omega_DFT}) to second order
in $1/R$ for spherically and cylindrically shaped liquid droplets.
For instance, the expansion of the density profile of the spherical
droplet reads:
\begin{equation}
\label{eq:expansion_rho}
\rho_s(r) =  \rho_0(z) + \frac{1}{R} \, \rho_{s,1}(z)
+ \frac{1}{R^2} \, \rho_{s,2}(z) + \ldots \,.
\end{equation}
The leading order correction to the density profile of the spherical interface
is twice that of the cylindrical interface, so it is convenient to define
$\rho_1(z) \!\equiv\! \rho_{s,1}(z) \!=\! 2 \, \rho_{c,1}(z)$ where $z \!\equiv\! r - R$.

The coefficients in the curvature expansion of the density are determined
from the curvature expansion of the Euler-Lagrange equation in Eq.(\ref{eq:EL_DFT}).
The result is that the (planar) density profile $\rho_0(z)$ is determined
from solving:
\begin{equation}
\label{eq:EL_0}
\mu_{\rm coex} = f^{\prime}_{\rm hs}(\rho_0) + \int \!\! d\vec{r}_{12} \; U_{\rm att}(r) \, \rho_0(z_2) \,,
\end{equation} 
and $\rho_1(z)$ follows from solving:
\begin{equation}
\label{eq:EL_1}
\mu_1 = f^{\prime\prime}_{\rm hs}(\rho_0) \, \rho_1(z_1) + \int \!\! d\vec{r}_{12} \; U_{\rm att}(r) \,
[ \, \rho_1(z_2) + \frac{r^2}{2} (1-s^2) \, \rho^{\prime}_0(z_2) \, ] \,,
\end{equation} 
where $\mu_1 \!=\! 2 \sigma / \Delta \rho$ \cite{Blokhuis93, Blokhuis06}.
For the evaluation of the curvature coefficients it turns out to be sufficient
to determine the density profiles $\rho_0(z)$ and $\rho_1(z)$ only.

Expressions for the curvature coefficients are now derived from the expansion of the
free energy in Eq.(\ref{eq:Omega_DFT}) to second order in $1/R$ and by comparing the
result to the expansions in Eq.(\ref{eq:sigma(R)}). This leads to the following two
equivalent expressions for the Tolman length \cite{Giessen98}:
\begin{equation}
\label{eq:delta_DFT_1}
\delta \sigma = \frac{1}{4} \int\limits_{-\infty}^{\infty} \!\!\! dz_1 \! \int \!\! d\vec{r}_{12} \;
U_{\rm att}(r) \, r^2 (1-s^2) \, \rho_0^{\prime}(z_1) \rho_1^{\prime}(z_2) \,,
\end{equation}
and
\begin{equation}
\label{eq:delta_DFT_2}
\delta \sigma = \frac{1}{4} \int\limits_{-\infty}^{\infty} \!\!\! dz_1 \! \int \!\! d\vec{r}_{12} \;
U_{\rm att}(r) \, r^2 (1-s^2) \, z_1 \, \rho_0^{\prime}(z_1) \rho_0^{\prime}(z_2)
- \frac{\mu_1}{2} \int\limits_{-\infty}^{\infty} \!\!\! dz \; z \, \rho_0^{\prime}(z) \,,
\end{equation}
The second order term in the expansion of the free energy for the {\em spherical}
interface, leads to the following expression for the combination of the rigidity
constants, $2k + \bar{k}$ \cite{Blokhuis13}:
\begin{eqnarray}
\label{eq:2k_kbar_DFT_1}
2k + \bar{k} &=& \frac{1}{4} \int\limits_{-\infty}^{\infty} \!\!\! dz_1 \! \int \!\! d\vec{r}_{12} \;
U_{\rm att}(r) \, r^2 (1-s^2) \, \rho_0^{\prime}(z_1) \rho_1(z_2) \\
&-& \frac{1}{4} \int\limits_{-\infty}^{\infty} \!\!\! dz_1 \! \int \!\! d\vec{r}_{12} \;
U_{\rm att}(r) \, r^2 (1-s^2) \, z_1^2 \, \rho_0^{\prime}(z_1) \rho_0^{\prime}(z_2) \nonumber \\
&+& \frac{1}{48} \int\limits_{-\infty}^{\infty} \!\!\! dz_1 \! \int \!\! d\vec{r}_{12} \;
U_{\rm att}(r) \, r^4 (1-s^4) \, \rho_0^{\prime}(z_1) \rho_0^{\prime}(z_2) \nonumber \\
&+& \int\limits_{-\infty}^{\infty} \!\!\! dz \left[ \frac{\mu_1}{2} z \, \rho_1^{\prime}(z)
+ \mu_1 \, z^2 \, \rho_0^{\prime}(z) + \mu_{s,2} \, z \, \rho_0^{\prime}(z) \right] \,, \nonumber
\end{eqnarray} 
where $\mu_{s,2} \!=\! - \sigma \, \Delta \rho_1 / (\Delta \rho)^2
- 2 \delta \sigma / \Delta \rho$ \cite{Blokhuis93, Blokhuis06} with
$\Delta \rho_1 \!\equiv\! \rho_{1,\ell} - \rho_{1,v}$. An alternative
expression that contains no reference to the chemical potential is
also derived in ref.~\cite{Blokhuis13}:
\begin{eqnarray}
\label{eq:2k_kbar_DFT_2}
2k + \bar{k} &=& - \frac{1}{2} \int\limits_{-\infty}^{\infty} \!\!\! dz_1 \! \int \!\! d\vec{r}_{12} \;
U_{\rm att}(r) \, r^2 (1-s^2) \, \rho_0^{\prime}(z_1) \rho_{s,2}^{\prime}(z_2) \\
&& - \frac{1}{4} \int\limits_{-\infty}^{\infty} \!\!\! dz_1 \! \int \!\! d\vec{r}_{12} \;
U_{\rm att}(r) \, r^2 (1-s^2) \, \rho_1^{\prime}(z_1) \rho_1^{\prime}(z_2) \nonumber \\
&& - \frac{1}{4} \int\limits_{-\infty}^{\infty} \!\!\! dz_1 \! \int \!\! d\vec{r}_{12} \;
U_{\rm att}(r) \, r^2 (1-s^2) \, z_1^2 \, \rho_0^{\prime}(z_1) \rho_0^{\prime}(z_2) \nonumber \\
&& + \frac{1}{48} \int\limits_{-\infty}^{\infty} \!\!\! dz_1 \! \int \!\! d\vec{r}_{12} \;
U_{\rm att}(r) \, r^4 (1-s^4) \, \rho_0^{\prime}(z_1) \rho_0^{\prime}(z_2) \,. \nonumber
\end{eqnarray}

To derive an expression for the bending rigidity $k$, an expansion
of the free energy to second order in $1/R$ is made for the {\em cylindrical}
interface. For the Tolman length the expressions in Eqs.(\ref{eq:delta_DFT_1})
and (\ref{eq:delta_DFT_2}) are recovered and one finds for the bending rigidity \cite{Blokhuis13}:
\begin{eqnarray}
\label{eq:k_DFT_1}
k &=& \frac{1}{8} \int\limits_{-\infty}^{\infty} \!\!\! dz_1 \! \int \!\! d\vec{r}_{12} \;
U_{\rm att}(r) \, r^2 (1-s^2) \, \rho_0^{\prime}(z_1) \rho_1(z_2) \\
&+& \frac{1}{64} \int\limits_{-\infty}^{\infty} \!\!\! dz_1 \! \int \!\! d\vec{r}_{12} \;
U_{\rm att}(r) \, r^4 (1-s^2)^2 \, \rho_0^{\prime}(z_1) \rho_0^{\prime}(z_2) \nonumber \\
&+& \int\limits_{-\infty}^{\infty} \!\!\! dz \left[ \frac{\mu_1}{4} z \, \rho_1^{\prime}(z)
+ \frac{\mu_1}{2} \, z^2 \, \rho_0^{\prime}(z) + 2 \mu_{c,2} \, z \, \rho_0^{\prime}(z) \right] \,, \nonumber
\end{eqnarray} 
where $\mu_{c,2} \!=\! - \sigma \, \Delta \rho_1 / (2 \, \Delta \rho)^2$
\cite{Blokhuis93, Blokhuis06}. As an alternative expression for $k$ that
contains no reference to the chemical potential is \cite{Blokhuis13}:
\begin{eqnarray}
\label{eq:k_DFT_2}
k &=& - \int\limits_{-\infty}^{\infty} \!\!\! dz_1 \! \int \!\! d\vec{r}_{12} \;
U_{\rm att}(r) \, r^2 (1-s^2) \, \rho_0^{\prime}(z_1) \rho_{c,2}^{\prime}(z_2) \\
&& - \frac{1}{8} \int\limits_{-\infty}^{\infty} \!\!\! dz_1 \! \int \!\! d\vec{r}_{12} \;
U_{\rm att}(r) \, r^2 (1-s^2) \, \rho_1^{\prime}(z_1) \rho_1^{\prime}(z_2) \nonumber \\
&& - \frac{1}{4} \int\limits_{-\infty}^{\infty} \!\!\! dz_1 \! \int \!\! d\vec{r}_{12} \;
U_{\rm att}(r) \, r^2 (1-s^2)^2 \, z_1^2 \, \rho_0^{\prime}(z_1) \rho_0^{\prime}(z_2) \nonumber \\
&& + \frac{1}{64} \int\limits_{-\infty}^{\infty} \!\!\! dz_1 \! \int \!\! d\vec{r}_{12} \;
U_{\rm att}(r) \, r^4 (1-s^2) (1+3s^2) \, \rho_0^{\prime}(z_1) \rho_0^{\prime}(z_2) \,. \nonumber
\end{eqnarray}
These expressions are derived in ref.~\cite{Blokhuis13} without reference
to any particular choice for the location of the dividing surface. In the
present analysis, it is important that the radius $R$ is defined according
to the equimolar radius in Eq.(\ref{eq:R}). Expansion of Eq.(\ref{eq:R})
to first order in $1/R$ then leads to the following two conditions for the
expanded profiles $\rho_0(z)$ and $\rho_1(z)$ \cite{Blokhuis93}:
\begin{equation}
\label{eq:equimolar}
\int\limits_{-\infty}^{\infty} \!\!\! dz \; z \, \rho_0^{\prime}(z) = 0 \,,
\hspace*{20pt} {\rm and} \hspace*{20pt}
\int\limits_{-\infty}^{\infty} \!\!\! dz \; z \, \rho_1^{\prime}(z) =
- \int\limits_{-\infty}^{\infty} \!\!\! dz \; z^2 \, \rho_0^{\prime}(z) \,.
\end{equation}
By making explicit use of this property for $\rho_0(z)$, Eq.(\ref{eq:delta_DFT_2})
may be rewritten as:
\begin{equation}
\label{eq:delta_DFT_3}
\delta \sigma = \frac{1}{4} \int\limits_{-\infty}^{\infty} \!\!\! dz_1 \! \int \!\! d\vec{r}_{12} \;
U_{\rm att}(r) \, r^2 (1-s^2) \, z_1 \, \rho_0^{\prime}(z_1) \rho_0^{\prime}(z_2) \,.
\end{equation}
Furthermore, using the properties for $\rho_0(z)$ and $\rho_1(z)$ in Eq.(\ref{eq:equimolar}),
the expression for $2k + \bar{k}$ in Eq.(\ref{eq:2k_kbar_DFT_1}) is rewritten as:
\begin{eqnarray}
\label{eq:2k_kbar_DFT_3}
2k + \bar{k} &=& - \frac{1}{4} \int\limits_{-\infty}^{\infty} \!\!\! dz_1 \! \int \!\! d\vec{r}_{12} \;
U_{\rm att}(r) \, r^2 (1-s^2) \, (z_1 + z_2) \, \rho_0^{\prime}(z_1) \rho_1^{\prime}(z_2) \\
&& + \frac{1}{48} \int\limits_{-\infty}^{\infty} \!\!\! dz_1 \! \int \!\! d\vec{r}_{12} \;
U_{\rm att}(r) \, r^4 (1-s^4) \, \rho_0^{\prime}(z_1) \rho_0^{\prime}(z_2) \,, \nonumber
\end{eqnarray}
while Eq.(\ref{eq:k_DFT_1}) is rewritten as:
\begin{eqnarray}
\label{eq:k_DFT_3}
k &=& - \frac{1}{8} \int\limits_{-\infty}^{\infty} \!\!\! dz_1 \! \int \!\! d\vec{r}_{12} \;
U_{\rm att}(r) \, r^2 (1-s^2) \, (z_1 + z_2) \, \rho_0^{\prime}(z_1) \rho_1^{\prime}(z_2) \\
&& + \frac{1}{8} \int\limits_{-\infty}^{\infty} \!\!\! dz_1 \! \int \!\! d\vec{r}_{12} \;
U_{\rm att}(r) \, r^2 (1-s^2) \, z_1^2 \, \rho_0^{\prime}(z_1) \rho_0^{\prime}(z_2) \nonumber \\
&& + \frac{1}{64} \int\limits_{-\infty}^{\infty} \!\!\! dz_1 \! \int \!\! d\vec{r}_{12} \;
U_{\rm att}(r) \, r^4 (1-s^2)^2 \, \rho_0^{\prime}(z_1) \rho_0^{\prime}(z_2) \,. \nonumber
\end{eqnarray}

\vskip 5pt
\noindent
\centerline{\rule{300pt}{1pt}}
\vskip 5pt
\noindent
Next, we verify that the expressions for $\delta$, $k$ and $2 k + \bar{k}$
as presented in Section \ref{sec-expressions}, reduce to the DFT expressions
listed above. To show the equivalence of the Triezenberg-Zwanzig and DFT
expression for the {\em surface tension of the planar interface}, one
approximates the direct correlation function by the attractive part of
the interaction potential \cite{Evans79, Evans90}:
\begin{equation}
\label{eq:C_0}
k_{\rm B} T \, C_0(z_1, z_2,r) \longrightarrow f^{\prime\prime}_{\rm hs}(\rho_0) \,
\delta(\vec{r}_2 - \vec{r}_1) + U_{\rm att}(r) \,.
\end{equation}
Inserting this approximation into the TZ expression for the surface tension
in Eq.(\ref{eq:sigma_TZ}) then immediately leads to the DFT expression in
Eq.(\ref{eq:sigma_DFT}).

Eq.(\ref{eq:C_0}) is easily generalized to {\em non-planar} geometries by
replacing $C_0$ and $\rho_0$ by the corresponding non-planar direct
correlation function and density profile. Since all the curvature expressions
feature a factor $r^2$ times the direct correlation, we may disregard
the delta-function contribution in Eq.(\ref{eq:C_0}). Using the fact
that the interaction potential is independent of $R$, we have the
following replacements:
\begin{eqnarray}
\label{eq:trans_1}
k_{\rm B} T \, \left[ \, \rho_s^{\prime} \, C_s \, \right]_1 &\longrightarrow&
\rho_1^{\prime}(z_1) \, U_{\rm att}(r) \,, \\
k_{\rm B} T \, \left[ \, \rho_s^{\prime} \rho_s^{\prime} \, C_s \, \right]_1 &\longrightarrow&
\left[ \, \rho_0^{\prime}(z_1) \rho_1^{\prime}(z_2) + \rho_1^{\prime}(z_1) \rho_0^{\prime}(z_2) \, \right] \, U_{\rm att}(r) \,, \nonumber \\
k_{\rm B} T \, \left[ \, \rho_s^{\prime} \rho_s^{\prime} \, C_s \, \right]_2 &\longrightarrow&
\left[ \, \rho_0^{\prime}(z_1) \rho_{s,2}^{\prime}(z_2) + \rho_1^{\prime}(z_1) \rho_1^{\prime}(z_2)
 + \rho_{s,2}^{\prime}(z_1) \rho_0^{\prime}(z_2) \, \right] \, U_{\rm att}(r) \,, \nonumber \\
k_{\rm B} T \, \left[ \, \rho_c^{\prime} \rho_c^{\prime} \, C_c \, \right]_2 &\longrightarrow&
\left[ \, \rho_0^{\prime}(z_1) \rho_{c,2}^{\prime}(z_2) + \rho_1^{\prime}(z_1) \rho_1^{\prime}(z_2) / 4
+ \rho_{c,2}^{\prime}(z_1) \rho_0^{\prime}(z_2)\, \right] \, U_{\rm att}(r) \,. \nonumber
\end{eqnarray}
Another property that is helpful in verifying that the expressions in Section
\ref{sec-expressions} reduce to the DFT expressions, is integration by parts
with respect to the parameter $s$. It is easily verified that:
\begin{eqnarray}
\label{eq:trans_2}
\int\limits_{-1}^{1} \!\! ds \; rs \, \rho_{\rm x}(z_2) &=&
\int\limits_{-1}^{1} \!\! ds \; \frac{r^2}{2} (1-s^2) \, \rho_{\rm x}^{\prime}(z_2) \,, \\
\int\limits_{-1}^{1} \!\! ds \; (1 - 3 s^2) \, \rho_{\rm x}(z_2) &=&
- \int\limits_{-1}^{1} \!\! ds \; rs \, (1-s^2) \, \rho_{\rm x}^{\prime}(z_2) \,. \nonumber
\end{eqnarray}
With the help of Eqs.(\ref{eq:trans_1}) and (\ref{eq:trans_2}), we may now show that:
(i), For the Tolman length $\delta$: Eq.(\ref{eq:delta_1}) reduces to Eq.(\ref{eq:delta_DFT_1})
and Eq.(\ref{eq:delta_2}) reduces to Eq.(\ref{eq:delta_DFT_3}); (ii), for the combination
$2 k + \bar{k}$: Eq.(\ref{eq:2k_kbar_1}) reduces to Eq.(\ref{eq:2k_kbar_DFT_2}) and
Eq.(\ref{eq:2k_kbar_2}) reduces to Eq.(\ref{eq:2k_kbar_DFT_3}); (iii), for the bending
rigidity $k$: Eq.(\ref{eq:k_1}) reduces to Eq.(\ref{eq:k_DFT_2})
and Eq.(\ref{eq:k_2}) reduces to Eq.(\ref{eq:k_DFT_3}).

\section{Equilibrium route -- Spherical droplet}
\label{app-sphere}

\noindent
In this appendix, we insert the expression for $\delta \rho_s(r)$ in
Eq.(\ref{eq:delta_rho_1}) into Eq.(\ref{eq:Laplace_s_3}) and expand
to order ${\cal O}(1/R^4)$. Three terms result that we shall investigate
separately:
\begin{eqnarray}
\label{eq:Laplace_s_4}
\frac{2 \sigma}{R^2} - \frac{4 \, \delta \sigma}{R^3} &=&
- k_{\rm B} T \int\limits_{0}^{\infty} \!\! dr_1 \int \!\! d\vec{r}_2 \;
C_s(r_1,r_2,r) \, \rho_s^{\prime}(r_1) \, \rho_s^{\prime}(r_2) \\
&& - \frac{k_{\rm B} T}{R^2} \int\limits_{0}^{\infty} \!\! dr_1 \int \!\! d\vec{r}_2 \;
C_s(r_1,r_2,r) \, \rho_s^{\prime}(r_1) \, \rho_1(r_2) \nonumber \\
&& - \frac{2 \, k_{\rm B} T}{R^3} \int\limits_{0}^{\infty} \!\! dr_1 \int \!\! d\vec{r}_2 \;
C_s(r_1,r_2,r) \, \rho_s^{\prime}(r_1) \, \rho_{s,2}(r_2) \nonumber \,.
\end{eqnarray}
The second Yvon-equation in Eq.(\ref{eq:Yvon_2_trans}) in spherical symmetry reads:
\begin{equation}
V_{\rm ext}^{\prime}(r_1) \, \hat{r}_1 = \int \!\! d\vec{r}_2 \; C_s(r_1,r_2,r) \, \rho_s^{\prime}(r_2) \, \hat{r}_2 \,.
\end{equation}
Considering the component in the $\hat{r}_1$ direction and considering
the limit of a vanishing external field, we have that:
\begin{equation}
0 = \int \!\! d\vec{r}_2 \; C_s(r_1,r_2,r) \, \rho_s^{\prime}(r_2) \, \frac{(r_1 + sr)}{r_2} \,,
\end{equation}
where we have used that $\hat{r}_1 \cdot \hat{r}_2 \!=\! (r_1 + sr) / r_2$.
This is conveniently rewritten as:
\begin{equation}
\label{eq:Yvon_2_s}
\int \!\! d\vec{r}_2 \; C_s(r_1,r_2,r) \, \rho_s^{\prime}(r_2) = 
\int \!\! d\vec{r}_2 \; C_s(r_1,r_2,r) \, \rho_s^{\prime}(r_2) \,
\left[ 1 - \frac{(r_1 + sr)}{r_2} \right] \,.
\end{equation}
The term in square brackets is expanded in the inverse radius to give:
\begin{eqnarray}
\label{eq:geometrical_term_1_s}
&& 1 - \frac{(r_1 + sr)}{r_2} = \frac{r^2}{2 R^2} (1-s^2) \nonumber \\
&& \hspace*{50pt} \times \, \left[ 1 - \frac{2}{R} (z_1 + sr) + \frac{3}{R^2}
\left( z_1^2 + 2 z_1 sr - \frac{r^2}{4} (1-5s^2) \right) + \ldots \right] \,,
\end{eqnarray}
where it is reminded that $z_1 \!\equiv\! r_1 - R$.
For the second and third term in Eq.(\ref{eq:Laplace_s_4}), we can use that
\begin{eqnarray}
\label{eq:2_3_term_s}
&& \int\limits_{0}^{\infty} \!\! dr_1 \int \!\! d\vec{r}_2 \;
C_s(r_1,r_2,r) \, \rho_s^{\prime}(r_1) \, \rho_{\rm x}(r_2) \nonumber \\
&& \hspace*{50pt} = \int\limits_{0}^{\infty} \!\! dr_1 \int \!\! d\vec{r}_2 \;
C_s(r_1,r_2,r) \, \rho_s^{\prime}(r_2) \, \rho_{\rm x}(r_1) \,
\left[ \frac{r_1^2}{r_2^2} - \frac{(r_1 + sr)}{r_2} \right] \,.
\end{eqnarray}
The first term in the square brackets results after making use
of the $1 \leftrightarrow 2$ symmetry of the direct correlation
function. The second term in the square brackets can be added
since Eq.(\ref{eq:Yvon_2_s}) indicates that it is equal to zero.
This addition is convenient since the sum of these two terms is
now of order ${\cal O}(1/R)$:
\begin{equation}
\label{eq:geometrical_term_2_s}
\frac{r_1^2}{r_2^2} - \frac{(r_1 + sr)}{r_2} = 
- \frac{2 sr}{R} + \frac{2 z_1 sr}{R^2} - \frac{r^2}{2 R^2} (1-7s^2) + \ldots \,.
\end{equation}
An important subtlety in the expansion of the direct correlation
is the fact that the second argument in $C_s(r_1,r_2,r)$ is expanded
around $r_2 \!=\! r_1 + rs$. The way this is done is explained in
ref.~\cite{Blokhuis92a} where the same mathematical technique is
used for the pair density. The result is that:
\begin{eqnarray}
\label{eq:C_s}
C_s(r_1,r_2,r) &=& \left[ 1 + \frac{r^2 (1-s^2)}{2 R} \, \frac{d}{r \, ds}
+ \frac{r^4 (1-s^2)^2}{8 R^2} \, \frac{d^2}{r^2 \, ds^2} \right. \\
&& \left. - \frac{r^2 (1-s^2) \, z_2}{2 R^2} \, \frac{d}{r \, ds} + \ldots \right] \, 
C_s(r_1,r_1 + rs,r) \,. \nonumber
\end{eqnarray}
A systematic expansion in $1/R$ of the right-hand-side in Eq.(\ref{eq:Laplace_s_4})
can now be made. After some algebra (using Eq.(\ref{eq:D4}) in Appendix \ref{app-extra})
and by comparing with the left-hand-side of Eq.(\ref{eq:Laplace_s_4}), we obtain the
TZ expression for $\sigma$ in Eq.(\ref{eq:sigma_TZ}) and find for the Tolman length
the following expression:
\begin{equation}
\label{eq:delta_s}
\delta \sigma = \frac{k_{\rm B} T}{8} \int\limits_{-\infty}^{\infty} \!\!\! dz_1 \! \int \! d\vec{r}_{12} \;
r^2 (1-s^2) \, \left[ \, \rho_s^{\prime}(z_1) \rho_s^{\prime}(z_2) \, C_s(z_1,z_2,r) \, \right]_1 \,.
\end{equation}
From the $1/R^4$ term in the expansion of the right-hand-side of Eq.(\ref{eq:Laplace_s_4}),
we have:
\begin{eqnarray}
\label{eq:2k_kbar_s}
& \left.
\begin{array}{ll}
0 = & - \frac{k_{\rm B} T}{12} \int\limits_{-\infty}^{\infty} \!\!\! dz_1 \! \int \! d\vec{r}_{12} \;
r^2 (1-s^2) \, \left[ \, \rho_s^{\prime}(z_1) \rho_s^{\prime}(z_2) \, C_s(z_1,z_2,r) \, \right]_2 \\
& - \frac{2 \, k_{\rm B} T}{3} \int\limits_{-\infty}^{\infty} \!\!\! dz_1 \! \int \! d\vec{r}_{12} \;
rs \, \rho_0^{\prime}(z_1) \rho_{s,2}(z_2) \, C_0(z_1,z_2,r) \\
& - \frac{k_{\rm B} T}{3} \int\limits_{-\infty}^{\infty} \!\!\! dz_1 \! \int \! d\vec{r}_{12} \;
rs \, \rho_1(z_2) \, \left[ \, \rho_s^{\prime}(z_1) \, C_s(z_1,z_2,r) \, \right]_1 \\
& - \frac{k_{\rm B} T}{4} \int\limits_{-\infty}^{\infty} \!\!\! dz_1 \! \int \! d\vec{r}_{12} \;
r^2 (1-s^2) \, z_1^2 \, \rho_0^{\prime}(z_1) \rho_0^{\prime}(z_2) \, C_0(z_1,z_2,r) \\
& + \frac{k_{\rm B} T}{48} \int\limits_{-\infty}^{\infty} \!\!\! dz_1 \! \int \! d\vec{r}_{12} \;
r^4 (1-s^4) \, \rho_0^{\prime}(z_1) \rho_0^{\prime}(z_2) \, C_0(z_1,z_2,r)
\end{array}
\hspace*{6pt} \right\} 2 k + \bar{k} \\
& \left.
\begin{array}{ll}
\hspace*{40pt} & + \frac{k_{\rm B} T}{6} \int\limits_{-\infty}^{\infty} \!\!\! dz_1 \! \int \! d\vec{r}_{12} \;
r (1-s^2) \, z_1 \, \left[ \, \rho_s^{\prime}(z_1) \rho_s^{\prime}(z_2) \, C_s(z_1,z_2,r) \, \right]_1 \\
& + \frac{k_{\rm B} T}{3} \int\limits_{-\infty}^{\infty} \!\!\! dz_1 \! \int \! d\vec{r}_{12} \;
rs \, z_1 \, \rho_0^{\prime}(z_1) \rho_1(z_2) \, C_0(z_1,z_2,r) \\
& - \frac{k_{\rm B} T}{12} \int\limits_{-\infty}^{\infty} \!\!\! dz_1 \! \int \! d\vec{r}_{12} \;
r^2 (1-3s^2) \, \rho_0^{\prime}(z_1) \rho_1(z_2) \, C_0(z_1,z_2,r) \\
& - \frac{k_{\rm B} T}{48} \int\limits_{-\infty}^{\infty} \!\!\! dz_1 \! \int \! d\vec{r}_{12} \;
r^4 (1-s^4) \, \rho_0^{\prime}(z_1) \rho_0^{\prime}(z_2) \, C_0(z_1,z_2,r)
\end{array}
\right\} - (2 k + \bar{k}) \nonumber 
\end{eqnarray}
The brackets indicate how the vanishing of this term may expressed as the sum of
two separate expressions for $2 k + \bar{k}$. This division may appear rather
arbitrary (the fifth and ninth term cancel, for instance), but is constructed
such that the separate contributions reduce to the DFT expressions for $2 k + \bar{k}$
as outlined in Appendix \ref{app-DFT}. This means that in principle a term could
be added and subtracted that has the property that it is zero when a mean-field
approximation is made, such as a term that contains $C_{s,1}(z_1,z_2,r)$ only.
Since none of the terms in Eq.(\ref{eq:2k_kbar_s}) have this property,
the presence of such a term is hard to imagine, however.

\section{Equilibrium route -- Cylindrical droplet}
\label{app-cylinder}

\noindent
In this appendix, we carry out the same analysis as in Appendix \ref{app-sphere}
but now for the {\em cylindrical} interface. Expanding $\rho_c(r)$ in $1/R$,
as in Eq.(\ref{eq:delta_rho_1}), gives
\begin{equation}
\delta \rho_c(r) = - \left[ \rho_c^{\prime}(r)
+ \frac{\rho_1(r)}{2 R^2} + \frac{2 \, \rho_{c,2}(r)}{R^3} + \ldots \right] \delta R \,.
\end{equation}
Inserting this expression for $\delta \rho_c(r)$ into Eq.(\ref{eq:Laplace_c_1})
and expanding to order ${\cal O}(1/R^4)$, we now have:
\begin{eqnarray}
\label{eq:Laplace_c_2}
\frac{\sigma}{R^2} - \frac{3 k}{2 R^4} &=&
- k_{\rm B} T \int\limits_{0}^{\infty} \!\! dr_1 \int \!\! d\vec{r}_2 \;
C_c(r_1,r_2,\varphi,r) \, \rho_c^{\prime}(r_1) \, \rho_c^{\prime}(r_2) \\
&& - \frac{k_{\rm B} T}{2 R^2} \int\limits_{0}^{\infty} \!\! dr_1 \int \!\! d\vec{r}_2 \;
C_c(r_1,r_2,\varphi,r) \, \rho_c^{\prime}(r_1) \, \rho_1(r_2) \nonumber \\
&& - \frac{2 \, k_{\rm B} T}{R^3} \int\limits_{0}^{\infty} \!\! dr_1 \int \!\! d\vec{r}_2 \;
C_c(r_1,r_2,\varphi,r) \, \rho_c^{\prime}(r_1) \, \rho_{c,2}(r_2) \nonumber \,.
\end{eqnarray}
In the absence of an external field, the second Yvon-equation in 
Eq.(\ref{eq:Yvon_2_trans}) in cylindrical geometry gives that:
\begin{equation}
\label{eq:Yvon_2_c}
\int \!\! d\vec{r}_2 \; C_c(r_1,r_2,\varphi,r) \, \rho_c^{\prime}(r_2) = 
\int \!\! d\vec{r}_2 \; C_c(r_1,r_2,\varphi,r) \, \rho_c^{\prime}(r_2) \,
\left[ 1 - \frac{(r_1 + sr)}{r_2} \right] \,.
\end{equation}
where we have used that $\vec{\nabla}_2 \, \rho_c(r_2) \!=\! \rho_c^{\prime}(r_2) \, \hat{r}_2$
and that $\hat{r}_1 \cdot \hat{r}_2 \!=\! (r_1 + sr) / r_2$. The term in
square brackets can be expanded in the inverse radius to give:
\begin{eqnarray}
\label{eq:geometrical_term_1_c}
&& 1 - \frac{(r_1 + sr)}{r_2} = \frac{r^2}{2 R^2} (1-s^2) \, \sin^2\!\varphi \\
&& \hspace*{50pt} \times \, \left[ 1 - \frac{2}{R} (z_1 + sr) + \frac{3}{R^2}
\left( z_1^2 + 2 z_1 sr - \frac{r^2}{4} (1-s^2) \, \sin^2\!\varphi + r^2 s^2 \right)
+ \ldots \right] \,. \nonumber
\end{eqnarray}
For the second and third term in Eq.(\ref{eq:Laplace_c_2}), we can use that
\begin{eqnarray}
\label{eq:2_3_term_c}
&& \int\limits_{0}^{\infty} \!\! dr_1 \int \!\! d\vec{r}_2 \;
C_c(r_1,r_2,\varphi,r) \, \rho_c^{\prime}(r_1) \, \rho_{\rm x}(r_2) \nonumber \\
&& \hspace*{50pt} = \int\limits_{0}^{\infty} \!\! dr_1 \int \!\! d\vec{r}_2 \;
C_c(r_1,r_2,\varphi,r) \, \rho_c^{\prime}(r_2) \, \rho_{\rm x}(r_1) \,
\left[ \frac{r_1}{r_2} - \frac{(r_1 + sr)}{r_2} \right] \,.
\end{eqnarray}
The first term in the square brackets results after making use
of the $1 \leftrightarrow 2$ symmetry of the direct correlation
function. The second term in the square brackets can be added
since Eq.(\ref{eq:Yvon_2_c}) indicates that it is equal to zero.
This addition is convenient since the sum of these two terms is
now of order ${\cal O}(1/R)$:
\begin{equation}
\label{eq:geometrical_term_2_c}
\frac{r_1}{r_2} - \frac{(r_1 + sr)}{r_2} = 
- \frac{sr}{R} + \frac{z_1 sr}{R^2} + \frac{r^2 s^2}{2 R^2} + \ldots \,.
\end{equation}
Analogous to Eq.(\ref{eq:C_s}), $C_c(r_1,r_2,\varphi,r)$ is expanded around
$r_2 \!=\! r_1 + rs$:
\begin{eqnarray}
\label{eq:C_c}
C_c(r_1,r_2,\varphi,r) &=& \left[ 1 + \frac{r^2 (1-s^2) \, \sin^2\!\varphi}{2 R} \, \frac{d}{r \, ds}
+ \frac{r^4 (1-s^2)^2 \, \sin^4\!\varphi}{8 R^2} \, \frac{d^2}{r^2 \, ds^2} \right. \\
&& \left. - \frac{r^2 (1-s^2) \, z_2 \, \sin^2\!\varphi}{2 R^2} \, \frac{d}{r \, ds} + \ldots \right] \, 
C_c(r_1,r_2 + rs,\varphi,r) \,. \nonumber
\end{eqnarray}
Again, a systematic expansion in $1/R$ of the right-hand-side in Eq.(\ref{eq:Laplace_c_2})
can now be made. After some algebra and by comparing to the left-hand-side
of Eq.(\ref{eq:Laplace_c_2}), we now obtain from the condition of the vanishing
of the first order term that:
\begin{eqnarray}
\label{eq:delta_c}
& \left.
\begin{array}{ll}
0 = & \frac{k_{\rm B} T}{4} \int\limits_{-\infty}^{\infty} \!\!\! dz_1 \! \int \! d\vec{r}_{12} \;
r^2 (1-s^2) \, z_1 \, \rho_0^{\prime}(z_1) \rho_0^{\prime}(z_2) \, C_0(z_1,z_2,r)
\end{array}
\hspace*{20pt} \right\} \delta \sigma \\
& \left.
\begin{array}{ll}
\hspace*{30pt} & - \frac{k_{\rm B} T}{8} \int\limits_{-\infty}^{\infty} \!\!\! dz_1 \! \int \! d\vec{r}_{12} \;
r^2 (1-s^2) \, \left[ \, \rho_s^{\prime}(z_1) \rho_s^{\prime}(z_2) \, C_s(z_1,z_2,r) \, \right]_1 \hspace*{8pt}
\end{array}
\right\} - \delta \sigma \nonumber 
\end{eqnarray}
where we have used the fact that to leading order $C_{c,1}(z_1,z_2,\varphi,r) \!=\!
\frac{1}{2} \, C_{s,1}(z_1,z_2,r)$. The $1/R^4$ term in Eq.(\ref{eq:Laplace_c_2}) now
supplies us with an expression for the bending rigidity:
\begin{eqnarray}
\label{eq:k_c}
& \left.
\begin{array}{ll}
k = & \frac{k_{\rm B} T}{3} \int\limits_{-\infty}^{\infty} \!\!\! dz_1 \! \int \! d\vec{r}_{12} \;
r^2 (1-s^2) \, \sin^2\!\varphi \, \left[ \, \rho_c^{\prime}(z_1) \rho_c^{\prime}(z_2) \, C_c(z_1,z_2,\varphi,r) \, \right]_2 \\
& + \frac{4 \, k_{\rm B} T}{3} \int\limits_{-\infty}^{\infty} \!\!\! dz_1 \! \int \! d\vec{r}_{12} \;
rs \, \rho_0^{\prime}(z_1) \rho_{c,2}(z_2) \, C_0(z_1,z_2,r) \\
& + \frac{k_{\rm B} T}{6} \int\limits_{-\infty}^{\infty} \!\!\! dz_1 \! \int \! d\vec{r}_{12} \;
rs \, \rho_1(z_2) \, \left[ \, \rho_s^{\prime}(z_1) \, C_s(z_1,z_2,r) \, \right]_1 \\
& + \frac{k_{\rm B} T}{4} \int\limits_{-\infty}^{\infty} \!\!\! dz_1 \! \int \! d\vec{r}_{12} \;
r^2 (1-s^2) \, z_1^2 \, \rho_0^{\prime}(z_1) \rho_0^{\prime}(z_2) \, C_0(z_1,z_2,r) \\
& - \frac{k_{\rm B} T}{64} \int\limits_{-\infty}^{\infty} \!\!\! dz_1 \! \int \! d\vec{r}_{12} \;
r^4 (1-s^2) (1+3s^2) \, \rho_0^{\prime}(z_1) \rho_0^{\prime}(z_2) \, C_0(z_1,z_2,r)
\end{array}
\right\} - k \hspace*{15pt} \\
& \left.
\begin{array}{ll}
& - \frac{k_{\rm B} T}{6} \int\limits_{-\infty}^{\infty} \!\!\! dz_1 \! \int \! d\vec{r}_{12} \;
r^2 (1-s^2) \, z_1 \, \left[ \, \rho_s^{\prime}(z_1) \rho_s^{\prime}(z_2) \, C_s(z_1,z_2,r) \, \right]_1 \\
& - \frac{k_{\rm B} T}{3} \int\limits_{-\infty}^{\infty} \!\!\! dz_1 \! \int \! d\vec{r}_{12} \;
rs \, z_1 \, \rho_0^{\prime}(z_1) \rho_1(z_2) \, C_0(z_1,z_2,r) \\
& + \frac{k_{\rm B} T}{12} \int\limits_{-\infty}^{\infty} \!\!\! dz_1 \! \int \! d\vec{r}_{12} \;
r^2 (1-3s^2) \, \rho_0^{\prime}(z_1) \rho_1(z_2) \, C_0(z_1,z_2,r) \\
& + \frac{k_{\rm B} T}{4} \int\limits_{-\infty}^{\infty} \!\!\! dz_1 \! \int \! d\vec{r}_{12} \;
r^2 (1-s^2) \, z_1^2 \, \rho_0^{\prime}(z_1) \rho_0^{\prime}(z_2) \, C_0(z_1,z_2,r) \\
& + \frac{k_{\rm B} T}{32} \int\limits_{-\infty}^{\infty} \!\!\! dz_1 \! \int \! d\vec{r}_{12} \;
r^4 (1-s^2)^2 \, \rho_0^{\prime}(z_1) \rho_0^{\prime}(z_2) \, C_0(z_1,z_2,r)
\end{array}
\hspace*{22pt} \right\} 2 k \nonumber 
\end{eqnarray}
Again, the brackets indicate how the bending rigidity may expressed as
the sum of two separate expressions. This division is constructed such
that the separate contributions both reduce to the DFT expressions in
Appendix \ref{app-DFT}, but, again, a term could be added and subtracted
that has the property that it is zero when a mean-field approximation
is made.

\section{Properties of direct correlation function}
\label{app-extra}

\noindent
In this appendix, we investigate some properties of the direct
correlation function of the planar interface, $C_0(z_1,z_2,r)$,
which turn out to be useful in rewriting the expression for
the Tolman length in Appendix \ref{app-sphere}.

In a uniform external field, the second Yvon equation in
Eq.(\ref{eq:Yvon_2_p}) reduces to:
\begin{equation}
\label{eq:D1}
0 = k_{\rm B} T \int \! d\vec{r}_{12} \; \rho_0^{\prime}(z_2) \, C_0(z_1,z_2,r) \,.
\end{equation}
The second Yvon equation in Eq.(\ref{eq:Yvon_2}) applied to the
situation in which a planar interface is curved by shifting the chemical
potential off-coexistence leads to:
\begin{equation}
\label{eq:D2}
\mu_1 = k_{\rm B} T \int \!\! d\vec{r}_{12} \;
[ \, \rho_1(z_2) + \frac{r^2}{2} (1-s^2) \, \rho^{\prime}_0(z_2) \, ] \, C_0(z_1,z_2,r) \,,
\end{equation} 
Eq.(\ref{eq:D1}) multiplied by $z_1 \, \rho_1(z_1)$ and integrated over $z_1$ gives:
\begin{equation}
\label{eq:D3}
0 =  k_{\rm B} T \int\limits_{-\infty}^{\infty} \!\!\! dz_1 \! \int \! d\vec{r}_{12} \;
(z_1 + rs) \, \rho_0^{\prime}(z_1) \rho_1(z_2) \, C_0(z_1,z_2,r) \,.
\end{equation}
Eq.(\ref{eq:D2}) multiplied by $z_1 \, \rho_0^{\prime}(z_1)$ and integrated over $z_1$ gives:
\begin{eqnarray}
\label{eq:D4}
&& \int\limits_{-\infty}^{\infty} \!\!\! dz_1 \! \int \! d\vec{r}_{12} \;
rs \, \rho_0^{\prime}(z_1) \rho_1(z_2) \, C_0(z_1,z_2,r) \\
&& \hspace*{20pt} =  \frac{1}{2} \int\limits_{-\infty}^{\infty} \!\!\! dz_1 \! \int \! d\vec{r}_{12} \;
r^2 (1-s^2) \, z_1 \, \rho_0^{\prime}(z_1) \rho_0^{\prime}(z_2) \, C_0(z_1,z_2,r) \,, \nonumber 
\end{eqnarray}
where Eq.(\ref{eq:D3}) has been used.


\begin{thebibliography}{0}

\bibitem{RW}
J.S. Rowlinson and B. Widom, {\em Molecular Theory of Capillarity}
(Clarendon, Oxford, 1982).

\bibitem{vdW}
J.D. van der Waals, Verhand. Kon. Akad. v Wetensch. Amst. Sect. 1, \textbf{8}, 1 (1893);
English translation in: J. Stat. Phys. \textbf{20}, 200 (1979).

\bibitem{Sullivan}
D.E. Sullivan, Phys. Rev. B {\bf 20}, 3991 (1979).

\bibitem{Evans79}
R. Evans, Adv. Phys. \textbf{28}, 144 (1979).

\bibitem{Evans90}
R. Evans, in {\em Liquids at Interfaces}, Les Houches XLVIII (1988),
eds. J. Charvolin, J.F Joanny, and J. Zinn-Justin (North-Holland, Amsterdam, 1990).

\bibitem{Evans84}
P. Tarazona and R. Evans, Mol. Phys. \textbf{52}, 847 (1984).

\bibitem{KB}
J.G. Kirkwood and F.P. Buff, J. Chem. Phys. \textbf{17}, 338 (1949).

\bibitem{Simulations1}
M.P. Allen and D.J. Tildesley, {\em Computer Simulation of Liquids},
(Oxford University Press, Oxford, 1987).

\bibitem{Simulations2}
D. Frenkel and B. Smit, {\em Understanding Molecular Simulation},
(Academic Press, San Diego, 1996).

\bibitem{Yvon}
J. Yvon, Proceedings of the IUPAP symposium on Thermodynamics, Brussels (1948).

\bibitem{TZ}
T.G. Triezenberg and R. Zwanzig, Phys. Rev. Lett. \textbf{28}, 1183 (1972). 

\bibitem{OZ}
L.S. Ornstein and F. Zernike, Proc. Acad. Sci. Amsterdam \textbf{17}, 793 (1914).

\bibitem{Tolman}
R.C. Tolman, J. Chem. Phys. \textbf{17}, 333 (1949).

\bibitem{Helfrich}
W. Helfrich, Z. Naturforsch. C \textbf{28}, 693 (1973).

\bibitem{Blokhuis92a}
E.M. Blokhuis and D. Bedeaux, Physica A \textbf{184}, 42 (1992).

\bibitem{Blokhuis93}
E.M. Blokhuis and D. Bedeaux, Mol. Phys. \textbf{80}, 705 (1993).

\bibitem{Giessen98}
A.E. van Giessen, E.M. Blokhuis, and D.J. Bukman, J. Chem. Phys. \textbf{108}, 1148 (1998).

\bibitem{Hemingway81}
S.J. Hemingway, J.R. Henderson, and J.S. Rowlinson, Faraday Symp. Chem. Soc. \textbf{16}, 33 (1981).

\bibitem{Henderson82}
J.R. Henderson and P. Schofield, Proc. R. Soc. Lond. A \textbf{380}, 211 (1982).

\bibitem{Schofield82}
P. Schofield and J.R. Henderson, Proc. R. Soc. Lond. A \textbf{379}, 231 (1982).

\bibitem{Lovett}
R. Lovett, P.W. DeHaven, J.J. Vieceli, and F.P. Buff, J. Chem. Phys. \textbf{58}, 1880 (1973).

\bibitem{Blokhuis99}
E.M. Blokhuis, J. Groenewold, and D. Bedeaux, Mol. Phys. \textbf{96}, 397 (1999).

\bibitem{Blokhuis09}
E.M. Blokhuis, J. Chem. Phys. \textbf{130}, 014706 (2009).

\bibitem{Blokhuis13}
E.M. Blokhuis and A.E. van Giessen, {\tt arXiv:cond-mat/13046557}.

\bibitem{Parry}
A.O. Parry and C.J. Boulter, J. Phys.: Condens. Matter \textbf{6}, 7199 (1994).

\bibitem{Meunier}
J. Meunier, J. Physique \textbf{48}, 1819 (1987).

\bibitem{Shape_equation}
Z. Ou-Yang and W. Helfrich, Phys. Rev. A \textbf{39}, 5280 (1989).

\bibitem{Gibbs}
J.W. Gibbs, {\em Collected works} (Dover, New York, 1961).

\bibitem{Barrett99}
J.C. Barrett, J. Chem. Phys. \textbf{111}, 5938 (1999).

\bibitem{Weeks1}
J.D. Weeks, {\it J. Chem. Phys.} \textbf{67}, 3106 (1977).

\bibitem{Weeks2}
J.D. Weeks and W. van Saarloos, J. Phys. Chem. \textbf{93}, 6969 (1989).

\bibitem{Weeks3}
J.D. Weeks, W. van Saarloos, D. Bedeaux, and E.M. Blokhuis, J. Chem. Phys. \textbf{91}, 6494 (1989).

\bibitem{TZ-KB}
M.H. Waldor and D.E. Wolf,, J. Chem. Phys. \textbf{85}, 6082 (1986).

\bibitem{FW}
M.P.A. Fisher and M. Wortis, Phys. Rev. B \textbf{29}, 6252 (1984).

\bibitem{Gompper}
G. Gompper and S. Zschocke, Phys. Rev. A \textbf{46}, 4386 (1992).

\bibitem{Blokhuis06}
E.M. Blokhuis and J. Kuipers, J. Chem. Phys. \textbf{124}, 074701 (2006).

\bibitem{FMT}
Y. Rosenfeld, Phys. Rev. Lett. \textbf{63}, 980 (1989);
R. Roth, J. Phys. Condens. Matter \textbf{22}, 063102 (2010).

\end{thebibliography}
\end{document}